\documentclass[reprint, amsmath,amssymb,
 aps,prx]{revtex4-2}
\usepackage{xcolor}
\usepackage[normalem]{ulem}
\usepackage{graphicx, color, braket}
\usepackage{dcolumn}
\usepackage{bm}
\usepackage[colorlinks=true,linkcolor=blue]{hyperref}

\renewcommand*{\eta}{\nu}

\begin{document}

\preprint{APS/123-QED}

\title{All-in-plane image sensors free from readout integrated circuits}

\author{Kirill Kapralov$^{1,2}$}
\email{kapralov.kn@phystech.edu}
\author{Ilya Mazurenko$^{1}$}
\author{Elizaveta Tarkaeva$^{3}$}
\author{Valentin Semkin$^1$}
\author{Oleg Kononenko$^{2,4}$}
\author{Maxim Knyazev$^{2,4}$}
\author{Viktor Matveev$^{2,4}$}
\author{Mikhail Kashchenko$^{1,2,5}$}
\author{Alexander Morozov$^{1,5}$}
\author{Ivan Domaratsky$^{1,2}$}
\author{Vladimir Kaydashev$^{6}$}
\author{Yana Litun$^{2}$}
\author{Aleksandr Kuntsevich$^{3}$}
\author{Alexey Bocharov$^{2}$}
\author{Dmitry Svintsov$^{1,2}$}%
\affiliation{$^1$Center for Photonics and 2d Materials, Moscow Institute of Physics and Technology (National Research University), Dolgoprudny 141700, Russia
}
\affiliation{$^{2}$Joint-Stock Company ''Skanda Rus'', Krasnogorsk 143403, Russia}
\affiliation{$^{3}$P.N. Lebedev Physical Institute of the Russian Academy of Sciences, 119991 Moscow, Russia}
\affiliation{$^{4}$Institute of Microelectronics Technology and High Purity Materials, Russian Academy of Sciences, Chernogolovka 142432, Russia}
\affiliation{$^{5}$Programmable Functional Materials Lab, Center for Neurophysics and Neuromorphic Technologies, Moscow 127495, Russia }
\affiliation{$^{6}$Laboratory of Nanomaterials, Southern Federal University, Rostov-on-Don 344090, Russia}

\date{\today}

\begin{abstract}
High resolution image sensors require electrical access to each individual pixel for signal readout. Such access is especially challenging for ultra-miniaturized pixels, for heterogeneously integrated sensing and readout layers in long-wavelength detectors, and for novel light-sensing materials with unestablished integration to silicon chips. Here, we introduce and experimentally validate a novel imaging approach that does not require electrical connections to individual pixels. The sensor matrix involves photoresistive pixels connected neighbor-to-neighbor and packed into a rectangular lattice. The signal readout is based on electrical impedance tomography applied to the photoresistance: the photovoltage is measured at the matrix boundary at various positions of injected bias current, and the image is reconstructed algorithmically. We present experimental validations for moderate-size infrared imagers based on multilayer graphene (24 pixels) and amorphous vanadium oxide (264 pixels). The reconstruction procedure is mathematically stable, sustainable to variations of pixel resistivity and photosensitivity, and its complexity is that of {\it linear} system solution. The proposed method enables unprecedented architecture simplification of imaging devices. 


\end{abstract}

\maketitle


Matrix photodetectors are fundamental optoelectronic devices with applications ranging from consumer electronics to medicine, industrial machine vision, and scientific research. The matrix imagers are currently available in multiple electromagnetic ranges, including infrared (IR) for thermal imaging and night vision, and terahertz for security inspection and defect sensing~\cite{Rogalski_progress_in_FPAs}. In long-wavelength devices, the photosensitive material has to be integrated heterogeneously with a silicon readout integrated circuit (ROIC). Integration  process requires ultra-precise wafer alignment, which is expensive, yield-limiting, and incompatible with emerging sensing materials. Even in monolithic image sensors based on silicon and its derivatives, integration of an amplifying transistor within each pixel introduces substantial technological complexity and high fabrication costs. The latter is dictated by multiple lithographic and doping steps, high-temperature annealing, and strict layer-to-layer alignment~\cite{ansari2024material}. A simpler alternative is represented by crossbar architectures~\cite{li2023electromagnetic}. Still, these suffer from signal crosstalk and parasitic capacitance~\cite{deng2012rram}. Eventually, crossbar image sensors can be implemented only with detectors possessing unidirectional diode-type current-voltage characteristics, a property being scarce for long-wavelength radiation sensors.

The complexity of ROICs and their integration with light-sensing layers is also a challenge for laboratory-stage studies of novel optoelectronic materials, such as graphene~\cite{Koppens_Graphene_image_sensor,Graphene_ROIC_integration}, metal dichalcogenides~\cite{PdTe2_image_sensor} and perovskites~\cite{Perovskite_Image_Sensors}. As a result, the research-grade matrix imagers based on emerging materials are based on custom electrode routing~\cite{Mennel_encoder}, while the number of pixels is limited by few tens. This prevents systematic scaling studies of new optoelectronic materials. A method enabling imaging without dedicated ROIC integration would unlock new opportunities for fundamental studies of photosensitive materials and device concepts.



Here, we introduce a method of imaging free from vertical integration of light-sensitive array with readout electronic circuits, thus removing the technological barriers to matrix scaling and material diversity. All the pixels in our architecture are located in a single plane, while the measurements of photoresponse are performed only at the perimeter of the matrix. Image reconstruction is enabled by the principles of electrical impedance tomography (EIT), a method initially developed for mapping of electrical conductivity in bulk media~\cite{adler2021electrical}. We apply the EIT principles to the reconstruction of distributed in-plane photoresistance, and thus retrieve the distribution of light intensity. 

\begin{figure*}[ht!]
\center{\includegraphics[width=1\linewidth]{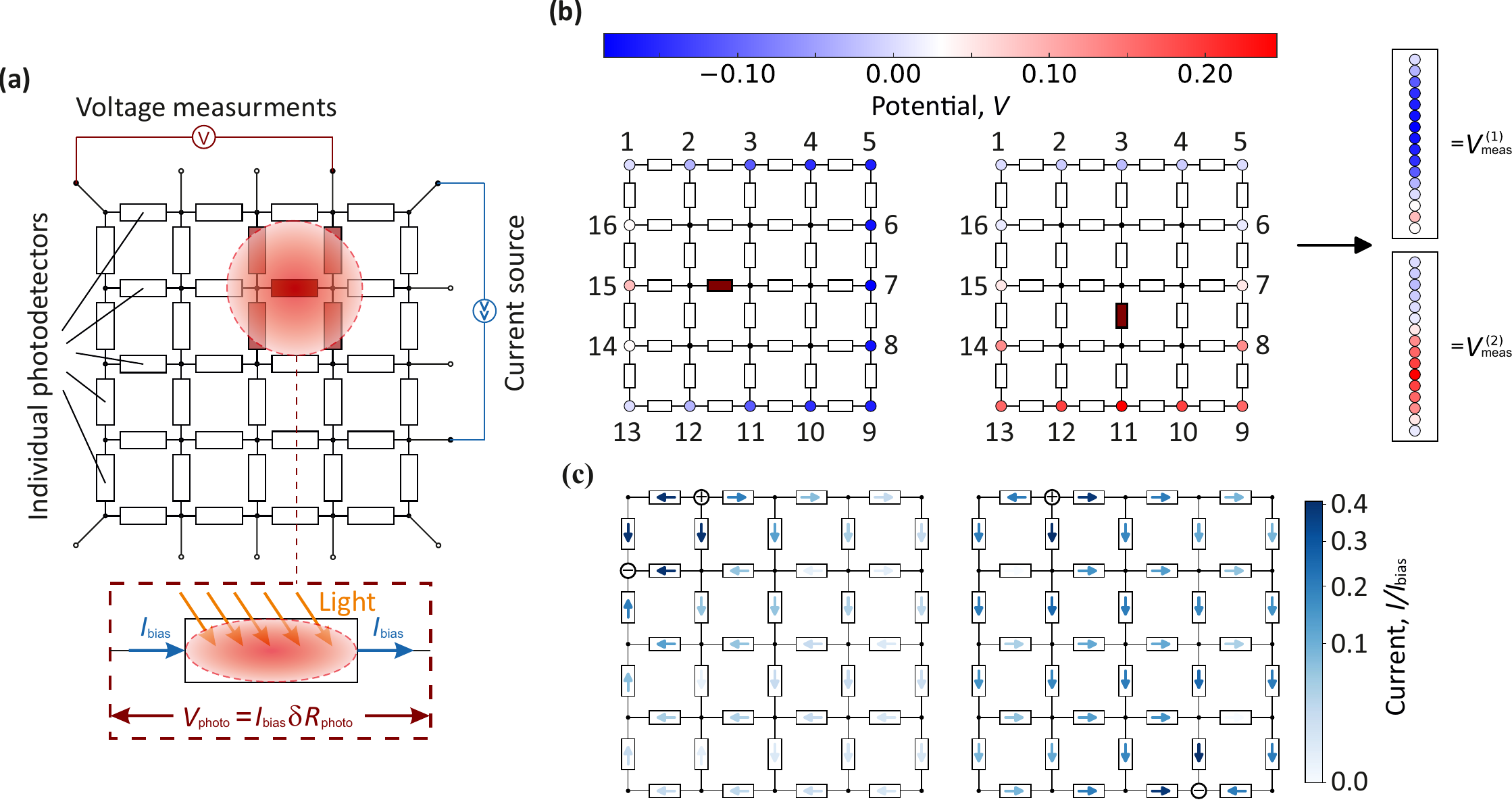} }
\caption{{\bf Principles of in-plane tomographic image sensor}.(a) Arrangement of photoresistive pixels comprising the image sensor and full signal acquisition scheme including peripheral electric contacts routed either to current source or voltmeter. (b) Sensitivity of the boundary photovoltage to the position of illumination spot: the spatial dependence of induced boundary potential (encoded with color) for two dissimilar positions of illuminated photoresistors, each mimiced as local voltage source with $\delta V({\bf r}_i)=1$ V  (c) Manipulation of spatial profile of detector sensitivity with bias current: profiles of local bias current $I({\bf r})$ for different injection points (marked with $+$ for source and $-$ for drain). Color intensity encodes the local current.}
\label{fig1}
\end{figure*} 

The EIT method relies on boundary voltage measurements at various positions of injected bias current. In original EIT, the measured voltage carries information on steady-state resistivity~\cite{brazey2022robust, adler2017electrical}; in our approach, the measured boundary voltage is associated with light and carries information on photoresistance. The photoresistance is generally small, as compared to dc resistance. As a result, the tomographic intensity reconstruction is a linear problem that does not face the challenges of high computational costs and spurious solutions inherent to conventional EIT~\cite{harrach2010exact, allers1991stability}. Our method can be referred to the field of \lq\lq reconstructive optoelectronics'', where the information about light properties is retrieved from multiple signals of tunable photodetectors~\cite{Computational_review_1,Deep_otpical_sensing}. The field has advanced greatly in spectrum~\cite{CompSpectr_review} and polarization~\cite{Deng_metasurface_array} retrieval with single pixels, and even in three-dimensional sensing with ROIC-integrated focal plane arrays~\cite{3D-Imaging_ROICs}. Still, it has not been applied to in-plane profiling of optical fields free from integrated circuitry.

\section*{Results}

\subsection*{Principles of in-plane tomographic image sensors} 

The designed tomographic image sensor is based on photoresistive detectors arranged in a plane square lattice with nearest-neighbor connections, as shown in Fig.~\ref{fig1} a. Unlike conventional pixel-addressed architectures, here the information about light intensity $J({\bf r})$, ${\bf r} = \{x,y\}$ is encoded non-locally in the potential distribution across the entire network. Electrical readout is performed only at the contacts along the matrix perimeter, as shown in Fig.~\ref{fig1} b. By sequentially applying the bias current between pairs of perimeter contacts, and measuring the resulting photovoltage between all other pairs, one obtains a large set of {\it nonlocal photoresistances}, each corresponding to a distinct current flow pattern through the matrix. These measurements form a dataset containing sufficient information to reconstruct the spatial distribution of photoresistance and light intensity.

The in-plane spatial resolution via boundary measurements only is guaranteed by two coexisting physical effects. First, application of bias current through different boundary points selectively activates the photosensitivity in different regions of the matrix. Each $i$-th photoresistive pixel positioned at ${\bf r}_i$ under illumination provides local photo-electromotive force $I({\bf r}_i)\,\delta\rho({\bf r}_i)$, where $\delta\rho({\bf r}_i)$ is the local photoresistance and $I({\bf r}_i)$ is the local bias current. Naturally, the distribution $I({\bf r}_i)$ depends on the choice of bias-injecting contacts. Two examples of such distributions are shown in Fig.~\ref{fig1} d, one activating the left upper edge of the matrix, and the other one activating the whole matrix in an almost uniform way. 

The second effect enabling spatial resolution is the dependence of boundary photovoltage pattern on the position of illuminated pixel. Two examples of such distinct patterns are shown in Fig.~\ref{fig1} c, where the highlighted pixel is assumed to generate unity photovoltage, and all other pixels assumed dark. The projected boundary voltage profiles $\delta V$ appear essentially dissimilar even for proximate active pixels.

Each of the mentioned effects is insufficient for full intensity reconstruction within the matrix plane, as the number of boundary points is well below the number of pixels $N_{\rm pix}$. Sufficient information is obtained if both current injection points are swept along the perimeter, and the respective sets of boundary voltages are recorded. Indeed, the matrix with $N$ pixels at each side contains $N_{\rm pix}=2N(N+1)$ photoresistors and $4N$ boundary electrical contacts. Having $N_e$ electrical connections at the perimeter ($N_e\le4N$) for current biasing and voltage sampling, we obtain $N_e(N_e-1)/2$ nonequivalent photoresistance measurements, where a factor of $1/2$ appears due to the current-voltage reciprocity. Complete image reconstruction is feasible if $N_e(N_e-1)/2\ge2N(N+1)$ or, for large matrices $N_e \ge 2N$. It implies that interrogation of only half of the boundary contacts is sufficient for complete image reconstruction, enabling compact and scalable imager architectures.

Passing to the practical implementation of tomographic imaging, we have derived a general relation between the local photoresistance $\delta\rho ({\bf r}_i)$, the distribution of bias current $I_\alpha({\bf r}_i)$ which is injected between a pair of contacts labeled by multi-index $\alpha$, and the boundary photovoltage $\delta V_{\alpha\beta}$ measured between nodes labeled by multi-index $\beta$:
\begin{equation}
\label{TellerTh}
\frac{\delta V_{\alpha\beta}}{I_\mathrm{bias}} =\sum_{i=1}^{N_{\rm pix}} 
\frac{I_\alpha({\bf r}_i)}{I_\mathrm{bias}} \frac{I_\beta({\bf r}_i)}{I_\mathrm{bias}}\,\delta\rho({\bf r}_i).
\end{equation}
In the linear regime, all pixels' photoresistances contribute to the boundary photovoltage independently, with scale factors $I_\alpha({\bf r}_i) I_\beta({\bf r}_i)/I^2_\mathrm{bias}$ that depend on both injection ($\alpha$) and measurement ($\beta$) points. Remarkably, the scale factor responsible for current-induced pixel activation $I_\alpha({\bf r}_i)/I_\mathrm{bias}$ (process in Fig.~\ref{fig1} d) is reciprocal to the scale factor of pixel-to-boundary signal propagation $I_\beta({\bf r}_i)/I_\mathrm{bias}$ (process in Fig.~\ref{fig1} c). This relation is guaranteed by Tellegen's theorem, and ensures the reciprocity of current and voltage probes.



Introducing the vector of measured non-local photoresistances $\delta {\boldsymbol{\rho}}_\mathrm{eff} =I_\mathrm{bias}^{-1} \{\delta V_{\alpha\beta}\}$, the vector of local photoresistances $\delta{\boldsymbol{\rho}} = \{\delta \rho({\bf r}_1),...\delta \rho({\bf r}_{N_{\rm pix}})\}$, and casting the tomographic equation (\ref{TellerTh}) in the matrix form, we obtain
\begin{equation}
\label{mainEq}
\hat{S}\,\delta{\boldsymbol{\rho}} = \delta \boldsymbol{\rho}_\mathrm{eff},
\end{equation}
where $\hat{S}$ is the dimensionless sensitivity matrix describing the electrical coupling between each detector and the boundary contacts. Equation~(\ref{mainEq}) represents the linearized form of impedance tomography for optoelectronic systems, where illumination induces small, spatially distributed perturbations of local light resistivity.

Image reconstruction amounts to the solution of (\ref{mainEq}) with respect to the photoresistance distribution $\delta {\boldsymbol{\rho}}$. The key step here is the accurate determination of the sensitivity matrix $\hat{S}$, for which two complementary approaches can be employed. First, the matrix $\hat{S}$ can be {\it computed} using the nominal values of detector resistances and direct Kirchhoff simulations of current spreading. In particular, if all resistances are identical, the matrix elements depend only on the matrix size (and topology, if non-trivial). In most practical cases, however, the pixel resistances are fluctuating randomly across the matrix. This brings us to the second, more rigorous approach. Here, absolute values of $\rho({\bf r}_i)$ are determined using {\it absolute} electrical impedance topography prior to the optical measurement. The absolute EIT is computationally demanding, still, it is performed only once at the calibration stage, and the computed $\hat{S}$-matrix is subsequently reused for multiple images.

\begin{figure*}[ht!]
\center{\includegraphics[width=1\linewidth]{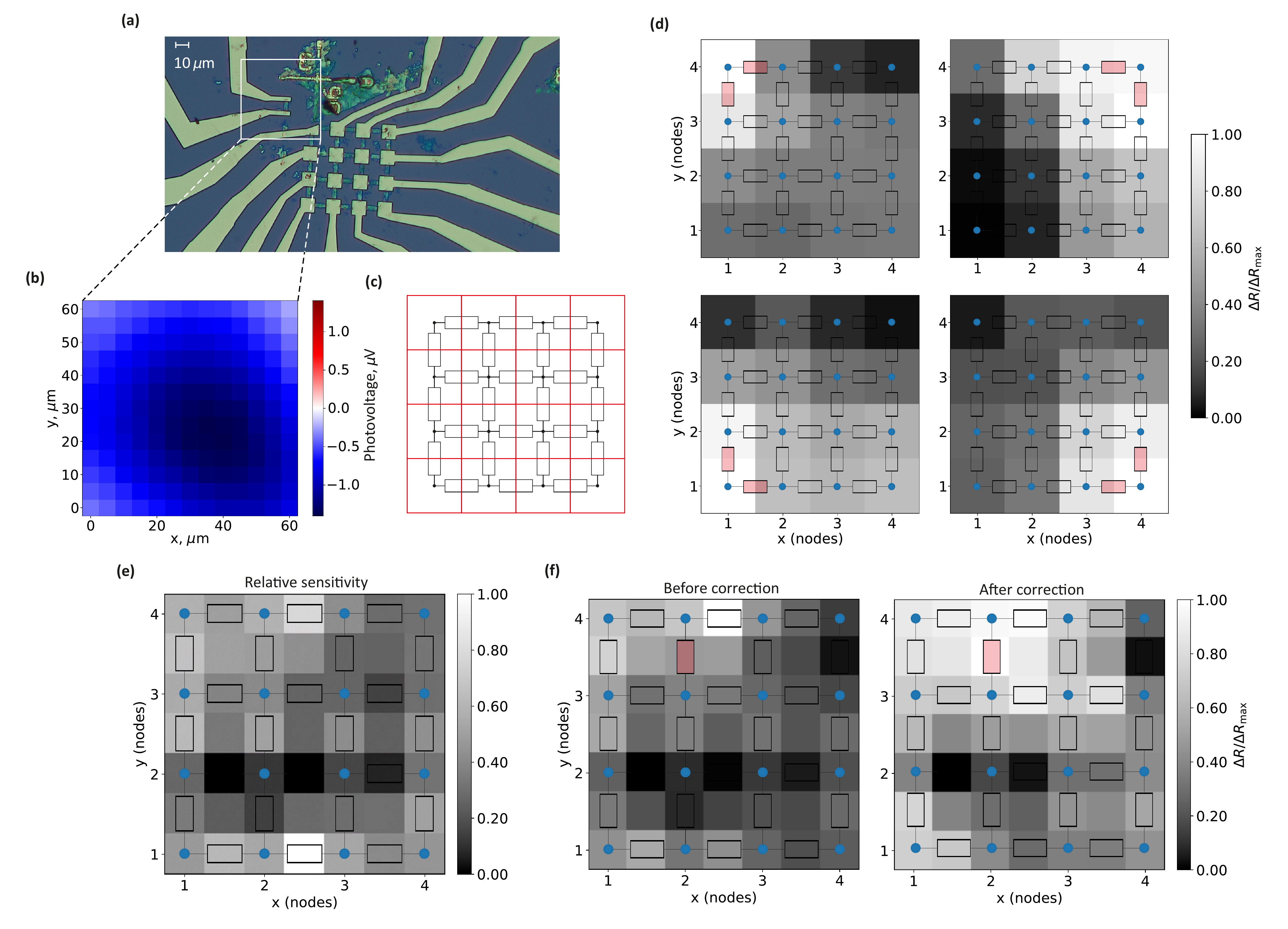} }
\caption{{\bf Proof-of-principle 24 pixel tomographic image sensor based on multilayer graphene} (a) Micro-photograph of matrix detector based on multilayer graphene. Individual reference pixel is enclosed in white square. (b) Map of photovoltage $\delta V = I_{\rm bias}\delta \rho$ recorded upon illumination of reference pixel with bias current $I_{\text{bias}}=20 \ \mu$A (c) Scheme for averaging  photo-resistance signals in a matrix photodetector for visualization. The detector signals in each square are to be averaged and displayed as pixel brightness (d) Reconstructed photoresistance in a 24-pixel graphene-based matrix detector when illuminated from various corners of the matrix. Photoresistors marked with red show the nominal position of illumination spot. (e) The relative sensitivity coefficients of the detectors in a $3 \times 3$ graphene matrix detector obtained as a result of applying the correction procedure based on permutation symmetries. (f) The result of the reconstruction of photoresistivity before and after applying the correction procedure when the red detector is illuminated. }
\label{fig3}
\end{figure*}

\subsection*{Experimental realization}

We proceed to the experimental validation of the in-plane tomographic imaging with moderate-size detector matrices in the technologically relevant mid-infrared range (wavelength $\lambda_0 = 8.6$ $\mu$m). The first matrix is based on multilayer graphene, has $N=3$ pixels at each side and $N_{\rm pix}=24$ pixels all together; electrical access is provided just to twelve pixels at the perimeter. The second matrix is based on amorphous vanadium oxide (VO$_x$) with $N=11$, $N_{\rm pix} = 264$ pixels overall, and $N_e=20$ electrical connections at the perimeter. Both materials are widely used in the realm of infrared detection~\cite{Wei_IR_graphene,Niklaus_MEMS_bolometers_review}, however, in the absence of thermal decoupling from substrate, their responsivity appears pretty low. To circumvent the problem, we use a quantum cascade laser with power $P \approx 10$ mW for array illumination, which results in relative photoresistance up to $\sim 1$ \% for beam focusing to diameter $d\approx 30-50$ $\mu$m.

Our proof of imaging concept relies on matrix illumination with focused laser beam in a pre-defined spot (set by motorized $xyz$-stage), and reconstruction of the spot image from interrogated boundary photovoltages at different injection currents. The on-demand connection of peripheral contacts either to current source or voltmeter (hereby lock-in amplifier) is achieved by in-house designed relay switchbox. The switchbox serves as an only external interface between the matrix detector and the measurement instruments; in an integrated design, equivalent functionality could be implemented directly on-chip.

To separate the photovoltage from the dc voltage drop, we used internal modulation of the laser source with frequency $f_{\rm las}$ ranging from tens of Hz to $\approx 900$ Hz, the particular frequency depending on resistance of IR-sensitive material and relevance of $RC$-delays. To get rid of photovoltages at the contacts between metal and sensing material (not related to photoconductivity), we also modulated  the bias current with low frequency $f_{\rm bias}$. The photoresistance signal was eventually measured the the difference frequency $|f_{\rm bias} - f_{\rm las}|$. 

\paragraph*{Graphene-based 24-pixel matrix detector.}
Our first device, the matrix with 24 pixels based on multilayer graphene (MLG), is shown in Fig.~\ref{fig3} (a). MLG with thickness $t\approx 15$ nm was grown by chemical vapor deposition on iron catalyst and wet-transferred onto Si/SiO$_2$ substrate, see \cite{brzhezinskaya2021engineering}
and Methods section for growth details. Pixels with dimensions $10$ $\mu$m $\times3$ $\mu$m were defined by electron lithography followed by plasma etching. A reference pixel was defined apart from the matrix [top left corner of Fig.~\ref{fig3} (a)] and used for control measurement of resistance and photoresistance. Its photovoltage map, obtained by $xy$-scan of the focused laser beam, is shown in Fig.~\ref{fig3} (b). With dark resistance $\rho_0\approx1.5$ k$\Omega$, the photovoltage signal at bias current $I_{\rm bias}=20$ $\mu$A is about 1 $\mu$V, corresponding to the photoresistance $\delta \rho\approx125$ m$\Omega$. Having checked the measurable value of pixel photoresistance, we proceed to the data acquisition for image reconstruction with a matrix detector.  

The resulting reconstructed images for laser illumination in four different corners of the matrix detector are shown in Fig.~\ref{fig3} (d). The intensity in the image is encoded by the reconstructed photoresistance, with white corresponding to the maximum measured $\delta \rho$, and black -- to zero signal. Assignment of color to a particular pixel is based on averaging of photoresistances belonging to a particular pixel, as explained schematically in Fig.~\ref{fig3} (d). We readily observe that our measurement scheme and reconstruction algorithm have clearly localized of the photoresponse in the illuminated quadrants with minimal cross-talk. In this setup with low pixel count, we have implemented the simplest form of linearized tomography, where the resistances of all pixels were assumed identical. The sensitivity matrix $\hat S$ can be computed analytically with such simplification. Despite this considerable simplification, the tomographic reconstruction has proved successful for minimal image retrievals.

We proceed to improve the reconstruction quality with the small-size matrix imager by implementing the correction for non-uniform pixel sensitivity. From global viewpoint, non-uniform pixel responsivity is ubiquitous in focal-plane arrays, where film thickness fluctuations, geometry variations, and local strain lead to inhomogeneous photoresponse. Therefore, the possibility for non-uniformity correction is of central importance for any image acquisition algorithm, including the proposed tomographic imaging. In this particular setup, the necessity for correction stems from inaccurate reconstruction for several spot positions. As example, illumination of a top vertically oriented pixel in the second column results in a bright reconstructed spot in the top row, as shown in Fig.~\ref{fig3} h. 


The implemented correction procedure relies on the translational symmetry of the detector lattice~\cite{Ratliff_translation_correction}. As the illumination spot is shifted by an integer number of pixels, the reconstructed pattern should shift by the same amount without amplitude variations, as explained in Fig.~\ref{fig3}e. Comparing the reconstructed images for laser positions displaced by a single pixel horizontally and vertically, one extracts the relative scaling factors for each row and column. The minimal information for complete responsivity correction can be obtained from only two such independent translations.


We have implemented the translation-based responsivity correction for graphene-based image sensor. The reconstructed spot after correction, Fig.~\ref{fig3} h, is positioned according to the nominal placement of laser beam, i.e. coincides with the top vertically oriented pixel in the second column. More interestingly, the correction procedure enables the evaluation of relative pixel responsivity variations across the matrix. This responsivity map, normalized by its maximum, is shown in Fig.~\ref{fig3} d. The root-mean-square scatter of the responsivity is $\sim 40$ \%, which represents quite a large variation. All the more, two of the pixels appear to have zero responsivity, probably due to the fracture of the contact. Despite all these technological imperfections, the tomographic algorithm provides stable image retrieval.

\paragraph*{264--pixel infrared image sensor based on amorphous vanadium oxide.}

\begin{figure*}[ht!]
\center{\includegraphics[width=1\linewidth]{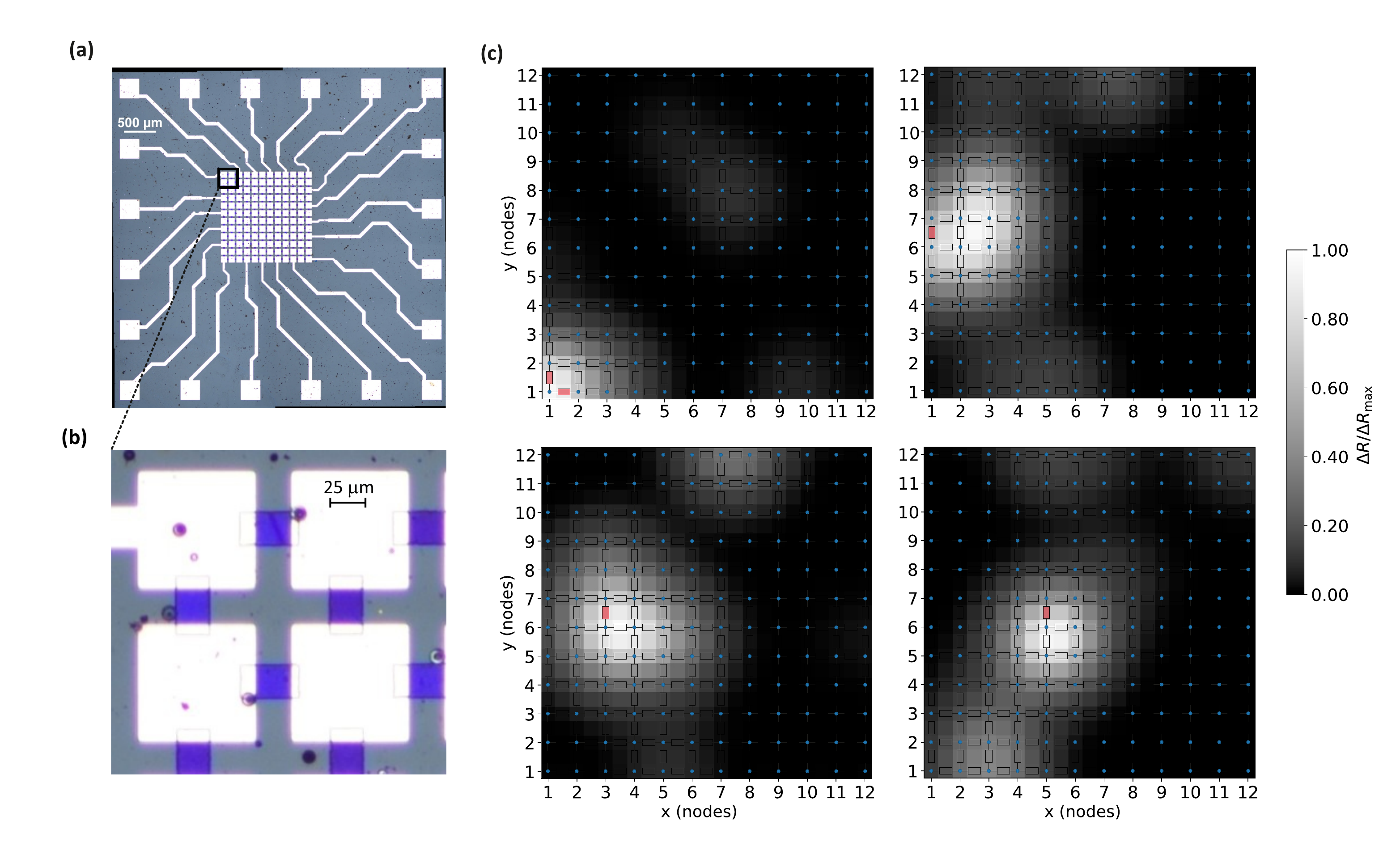} }
\caption{{\bf Larger-scale tomographic imager based on vanadium oxide.} (a)-(b) Optical micro-photograph of the in-plane matrix detector based on amorphous VO$_x$ with 264 pixels.(c) Reconstructed photoresistance distributions $\delta \rho({\bf r})$ for illumination at red detectors of the matrix, demonstrating stable localization of the illumination spot using tomographic procedure.}
\label{fig2}
\end{figure*}

Proceeding to the larger-resolution image sensors, we study the second device with 264 photoresistors based on amorphous VO$_x$, Fig.~\ref{fig2} (a). The IR-sensitive films were grown by e-beam evaporation of vanadium in the oxygen atmosphere, see \cite{tarkaeva2024} and Methods for details. The matrix has $N=11$ pixels at each side and $N_e=20$ active electrodes available on the periphery. Each pixel has dc resistance $\rho\sim 150$ k$\Omega$, its photoresistance upon focused illumination is $\delta \rho\sim 450\,\Omega$, which reaches $\sim0.3$ \% of the total resistance. Relatively large photoresistance is ensured by high thermal coefficient of resistance TCR$\approx -1.4$ \%/K of vanadium oxide films, measured independently from optoelectronic studies.

In larger-size tomographic image sensors, accurate determination of the sensitivity matrix $\hat{S}$ is crucial. In turn, it relies on the knowledge of static resistances for all pixels in the absence of illumination. This resistance mapping was performed by first implementing the dc electrical impedance tomography in the absence of illumination. 

After the resistance calibration, we implemented the reconstruction of photoresistances for several positions of illumination spot. These positions are marked with red in Fig.~\ref{fig2} c. One readily observes that our reconstruction algorithm has again succeeded in identifying the spot position. The reconstructed maps of photoresistance are shown in Fig.~\ref{fig2} c; the assignment of color to a particular pixel is based on the photoresistances averaged according to the ''chessboard'' pattern, shown in Fig.~\ref{fig2} (b). The localization appeared equally efficient for illumination near the matrix edges and its center. The latter case is more challenging due to the low values of bias current density in the matrix center and, hence, due to the relatively low sensitivity of the central pixels. 

Artifacts in the reconstructed images, seen as secondary spots with lower photoresistance, may have appeared due to the limited number of electrical contacts at the matrix perimeter. These enable only $N_e(N_e-1)/2=190$ independent electrical measurements, which is below the total number of pixels $N_{\rm pix}=264$. For this reason, the resistors were artificially clustered in groups when solving both absolute and differential tomography problems. A highly positive outcome of this study is the possibility of identifying spot positions with very limited number of electrical measurements.

\section*{Discussion}

The demonstrated planar detector architecture based on tomographic reconstruction represents an attractive paradigm in optoelectronic imaging. Unlike conventional pixel-addressable arrays, where each element is individually read out through multilayer interconnects and active circuitry, here optical information is encoded collectively in a resistive response at the periphery of a single planar conductive sheet. This removes the need for transistor integration and vertical metallization, offering a fabrication-tolerant and geometrically scalable alternative to CMOS-based matrices.

A definitive advantage of the approach is its independence from the microscopic mechanism of photoresistance. The underlying effects can include bolometric heating (relevant to silicon and vanadium oxide infrared imagers~\cite{Niklaus_MEMS_bolometers_review}), interband electron-hole excitation (relevant to mercury cadmium telluride~\cite{HgTe_photoconductivity} and 2d black phoshorus~\cite{Interband_PC_blackP}), and photogating~\cite{Photogating_low_dimensional}, to name a few. The sensitive material need not be two-dimensional, as we have demonstrated with the example of vanadium oxide films. In fact, a similar image plane can be realized with bulk photoresistors (e.g., made of cadmium sulfide) arranged into a rectangular lattice. The only limitation to the detection mechanism is that radiation should alter the material resistance. In this regard, zero-bias photovoltaic effect (e.g., in a $p-n$ or Schottky junction) does not fit the proposed architecture. The reason is that variations in bias current $I_\alpha({\bf r}_i)$ that reconfigure the spatial sensitivity of photoresistive plane, would not affect the zero-bias detection, least in the linear mode. Similar imaging functionality can, in principle, be achieved with gated zero-bias detectors~\cite{3D-Imaging_ROICs}, still, it requires a separate study. 

The number of electrical contacts required for full image reconstruction $N_e$ equals approximately to the half of boundary nodes, and scales as square root of total pixel count $N_e \approx \sqrt{2N_{\rm pix}}$. The scaling ensures that even large-format matrices can be read out through a limited number of electrical connections. We may note that the square-root scaling is similar to that for crossbar architectures. Continuing the comparison of in-plane tomographic and crossbar schemes, we note that the problem of shunt currents inherent to crossbars becomes the operating principle of tomographic sensors. Instead of trying to avoid the current spreading, here we manipulate the spreading in a controllable way, and use the known spread patterns for image reconstruction. From technology viewpoint, the supremacy of tomographic sensors compared to crossbars is the complete absence of vertical bar overlaps that are prone to dielectric breakdowns.

In our proof-of-concept demonstration, measurements of photovoltage at each injection current were performed sequentially, which resulted in large timescales of image acquisition. In advanced designs, it is possible to decouple the current-injection and voltage-readout contacts, and read out all signal voltages in parallel. The number of electrical permutations for bias current injection would scale as $N_{\rm pix}^{1/2}$, further improving the response times. We can also expect that the proposed architecture can operate as in-sensor classifier~\cite{Mennel_encoder,Wearable_reservoir_computing} by providing unique boundary voltage patterns $\delta V$ for selected recognized images. Learning of such inference sensor would amount to the selection of bias current profile $I_\alpha({\bf r})$.


Beyond imaging, the tomographic approach opens new directions for fundamental research. It is possible to generalize the photoresistance reconstruction algorithm to continuous thin-film materials with multiple electrical contacts, the most recognized example being the 2d Hall bar. As a result, information about distributed local photoresistance and local field hot spots under illumination can be extracted with spatial resolution order of film size divided by the number of contacts squared. Previously, such information was attainable only with advanced near-field microscopy techniques~\cite{Local_PC_3,Local_PC_4}. We may say that photoresistance tomography can become a kind of microcopy without a microscope, and provide information about local electromagnetic absorption by smart processing of multiple photovoltage measurements.



\section*{Methods}

\subsection*{Tomographic image reconstruction}

We start by proving the fundamental tomographic equation (\ref{TellerTh}) relating the photoresistances $\delta\rho({\bf r}_i)$ with boundary potentials $\delta V_{\alpha\beta}$. Each photoresistor is equivalent to the voltage source with electromotive force $\delta {\mathcal E}_\alpha({\bf r}_i) = I_\alpha({\bf r}_i)\delta\rho({\bf r}_i)$. Linearity of the circuit ensures that photovoltages produced by individual pixels add up independently. To find the peripheral voltage $\delta V_{\alpha\beta}$ induced by $\delta {\mathcal E}({\bf r}_i)$, we use Tellegen's reciprocity theorem. Namely, if an EMF source placed in the $k$-th circuit branch induces current $I_n$ in the $n$-th circuit branch, then the same source placed in the $n$-th branch would induce the same current $I_k = I_n$ in the $k$-th branch. Now it follows that $\delta V_{\alpha\beta} = e_\beta \delta {\mathcal E}_\alpha({\bf r}_i)$, with the weightening factor $e_\beta = I_\beta({\bf r}_i)/I_{\rm bias}$. While the current $I_\alpha$ is truly injected between contact pair $\alpha$, the current $I_\beta$ is fictitious. It would appear in the circuit if the same current source is connected in place of voltmeter at the terminal pair $\beta$. Such relation between $\delta V_{\alpha\beta}$ and $\delta\rho({\bf r}_i)$ allows simple construction of the sensitivity matrix $\hat{S}$.

To reconstruct $\delta\rho({\bf r}_i)$ we need to solve the linear equation~(\ref{mainEq}) with the matrix $\hat{S}$, which is inherently ill-conditioned, reflecting the diffusive nature of current propagation in a resistive sheet. To obtain physically meaningful reconstructions, we constrain the solution space by assuming that illumination induces resistivity changes of uniform sign across the matrix detector. Under this constraint, the solution is obtained using a non-negative least-squares (NNLS) approach, which provides natural regularization and suppresses noise amplification for ill-conditioned high-dimensional matrices~\cite{slawski2013non}.

For the VO$_x$ matrix the number of resistive elements ($N_{pix}=264$) substantially exceeds the number of independent electrical measurements that can be performed with only $N_e=20$ electrical connections on the periphery. Direct inversion of equation~(\ref{mainEq}) with $N_{pix}$ unknowns is therefore strongly undetermined and highly sensitive to noise. To reduce the dimensionality of the problem while preserving spatial localization of the reconstructed response, we expand the photoresistivity change $\delta\rho(\mathbf{r})$ in a reduced basis of smooth Gaussian functions:
\begin{equation}
\delta\rho(\mathbf{r}_i) = \sum_{k=1}^{N_b} a_k\,g_k(\mathbf{r}_i),
\label{eq:gauss_expansion}
\end{equation}
where $g_k(\mathbf{r})$ are normalized Gaussian basis functions and $a_k\geq 0$ are their amplitudes. In this work we choose a regular $6\times 6$ grid of basis functions ($N_b=36$) with the centers $\mathbf{R}_k$:
\begin{equation}
g_k(\mathbf{r}) =
\frac{1}{2\pi\sigma^2}
\exp\!\left(
-\frac{|\mathbf{r}-\mathbf{R}_k|^2}{2\sigma^2}
\right),
\label{eq:gauss_def}
\end{equation}
where the width $\sigma$ is chosen to be comparable to half the distance between adjacent peaks of the Gaussian basis functions.

\subsection*{Device fabrication}

Graphene-based matrix detectors were fabricated using optical and electron-beam lithography combined with oxygen plasma etching and metal deposition. Multilayer graphene films were grown on iron catalysts using the low-pressure chemical vapor deposition (LPCVD) method with a single injection of acetylene~\cite{brzhezinskaya2021engineering}, and transferred from iron onto pre-cleaned Si/SiO$_2$ substrates using PMMA-assisted wet transfer. The thickness of the synthesized films ranged from 20 to 40~nm. Prior to patterning, the substrates were annealed in argon at 350~$^\circ$C for 1~h under 0.05~MPa pressure to improve adhesion. A two-layer PMMA resist stack (495K~A4/950K~A5) was spin-coated at 3000–4000~rpm and baked at 170~$^\circ$C. The pattern of detector elements and electrode geometry was defined by electron-beam lithography (Crestec~CABL-9000C) with exposure doses of 450–600~$\mu$C~cm$^{-2}$, depending on the feature size. The exposed regions were developed in an isopropanol:water mixture (3:1) for 60~s. Reactive-ion etching of graphene was performed in a COVANCE plasma treatment unit using oxygen plasma at 150~W for 20~min. A MgO mask was used to protect the active regions. 
Metallic contacts were formed by magnetron sputtering of gold (60~nm) with an adhesion layer of titanium (5~nm). 
Magnetron sputtering was chosen over e-beam evaporation to ensure conformal coating of graphene edges and reliable electrical contact. After liftoff in 
NMP, the devices were rinsed in isopropanol and dried in nitrogen.


VO$_x$-based matrix detector was fabricated on a sapphire (Al$_2$O$_3$) substrate using optical lithography and thin-film deposition techniques. First, the pattern for electrodes and detector elements was defined in a photoresist stack. A LOR 7A sublayer was spin-coated at 4000 rpm and baked at 160$^\circ$C for 3 minutes, followed by an AZ 1512 HS photoresist layer spin-coated at 3000 rpm and baked at 115$^\circ$C for 30 seconds. This stack was patterned using a Heidelberg $\mu$PG-101  direct laser writing system. A titanium/aluminum layer (20 nm/50 nm) was deposited using electron-beam evaporation (Plassys MEB-550) at a rate of 0.1 nm/s, followed by a lift-off process. The next photoresist layer was then coated with the same parameters, and a pattern for the amorphous vanadium oxide detector elements was defined using the same lithographer. The VO$_x$ sensing layer was deposited by reactive electron-beam evaporation from a vanadium source in an oxygen atmosphere (0.3 sccm flow) at a rate of 0.05 nm/s and at room temperature~\cite{tarkaeva2024}. The second lift-off was then performed.
The resulting VO$_x$ film has a thickness of approximately 170 nm and a temperature coefficient of resistivity of about -1.4\%/K.

\subsection*{Optoelectronic characterization}

Optoelectronic measurements were performed using a 8.6 $\mu$m quantum cascade laser (QCL) as a radiation source, with an output power $P \approx 10$ mW. The subsequent optical system consisted of a quarter-wave plate, a polarizer, and a focusing lens. This system was used to set the polarization along the matrix diagonal, ensuring equivalent pixel sensitivity in columns and rows.
The QCL with the optical system were mounted on an $xyz$-stage, enabling precise positioning with respect to the sample and automated alignment. Through the ZnSe window of the vacuum chamber, the beam was focused onto the sample into a spot with diameter of $\sigma$~=~20~$\mu$m.



To single out the contribution to the voltage signal resulting from photoresistance, a double lock-in detection scheme was employed. The laser was modulated with frequency $f_\mathrm{las}$ set by internal modulation of drive current, while the bias current was modulated with frequency $f_\mathrm{bias}$. The signal at the difference frequency $|f_\mathrm{bias}-f_\mathrm{las}|$ carries information about photoresistance, while dc voltage drops and photovoltages at metal contacts not related to photoconductivity are excluded.
The lock-in amplifier SR 860 was used both as a voltage measurement unit and current source, in the latter case its voltage output was connected through a large resistor with $R_{\rm load} \approx 1$ M$\Omega$ to the peripheral contact of the measured matrix. 

An in-house designed relay switchbox was used to route the peripheral electrical contacts of the matrix detector between the current source, the voltage input of lock-in amplifier, or the 'floating' (disconnected) state. The switchbox has a rectangular structure where rows of "horizontal" wires intersect with columns of "vertical" wires. At each intersection, a solid-state photorelay is placed. Each of these relays is independently switched on or off, controlled from the PC through intermediate digital logic. The system is shown schematically in the figure 4.



Optoelectronic measurements aimed at image reconstruction were carried out in two steps. First, the $xy$-stage position $(x_0,~y_0)$ at which the laser spot illuminates the reference pixel in the matrix corner was determined. To this end, the spatial map of the photosignal was measured while the readout scheme was connected to the reference pixel pads. The photosignal spot center on the map was taken as the reference pixel position. Second, the laser spot was placed in an arbitrary location $(x,~y)$ on the matrix, and a full set of boundary measurements was acquired. The set was subsequently used as an input for the tomographic reconstruction procedure governed by Eq.~\ref{TellerTh}.

\section*{Acknowledgments}

The work was supported by the Ministry of Science and Higher Education of the Russian Federation (FSMG-2025-0005). K.K was supported by the grant No. 25-29-20273 of the Russian Science Foundation (development and implementation of the tomographic image reconstruction method).
E.T. was supported by Basic Research Program of the HSE University (development of the technology and fabrication of VO$_x$ samples), VO$_x$ samples were fabricated using the Shared Facility Center of the P.N. Lebedev Physical institute. O.K., M.K., V.M. acknowledge funding from the Ministry of Science and Higher Education of the Russian Federation (state task No. 075-00296-26-00) (development of the technology and fabrication of graphene samples).

\section*{Competing interests}
The authors declare no competing interests.



\bibliography{references}

\end{document}